\documentclass{MEM}
\usepackage{natbib}\usepackage{txfonts}\usepackage{balance}
\usepackage{graphicx}
\usepackage[a4paper,breaklinks,dvipdfm]{hyperref}
\idline{84}{1}
\usepackage{txfonts} 

\newcommand{\AGILE}{{\sl AGILE}}

\newcommand{\cha}{{\sl Chandra}}
\newcommand{\Fermi}{{\sl Fermi}}
\newcommand{\GeV}{${\rm Ge\!V}$}
\newcommand{\keV}{${\rm ke\!V}$}
\newcommand{\MeV}{${\rm Me\!V}$}

\begin{document}
\def\teff{$T\rm_{eff }$}
\def\kms{$\mathrm {km s}^{-1}$}

\title{
Monitoring the Crab Nebula with Chandra
}

   \subtitle{A Search for the Location of the $\gamma$-Ray Flares}

\author{
Martin C. \,Weisskopf\inst{1} 
          }

  \offprints{Martin C. Weisskopf}
 
\institute{NASA/Marshall Space Flight Center, Astrophysics Office, Huntsville AL 35812, USA\email{martin.c.weisskopf@nasa.gov}}

\authorrunning{Weisskopf }

\titlerunning{Monitoring of the Crab}

\abstract{
Subsequent to announcements by the \AGILE~ and by the \Fermi-LAT teams of the discovery of $\gamma$-ray flares from the Crab Nebula in the fall of 2010, an international collaboration has been monitoring X-Ray emission from the Crab on a regular basis using the \cha~ X-Ray Observatory. Observations occur typically once per month when viewing constraints allow. The aim of the program is to characterize in depth the X-Ray variations within the Nebula, and, if possible, to much more precisely locate the origin of the $\gamma$-ray flares. In 2011 April we triggered a set of \cha~ Target-of-Opportunity observations in conjunction with the brightest $\gamma$-ray flare yet observed. We briefly summarize the April X-ray observations and the information we have gleaned to date.
}

\maketitle{}

\section{Introduction}
Since 2007, the \AGILE~ and \Fermi~ satellites have detected several $\gamma$-ray flares in the $0.1-1$~\GeV\ range from the Crab Nebula \citep{tav11,abd11,str11a,bue12}.  
The largest flares exhibit variability on timescales as short as hours. 
Prior to the 2011-April flare, the only Crab $\gamma$-ray flare covered by a multi-wavelength observing program was the 2010-September flare, which triggered only post-flare observations in radio, optical, and X-ray bands.
Despite the brightness of the $\gamma$-ray flares, there has been no clear evidence for correlated variations in radio \citep{lob11,wei12}, near-infrared \citep{kan10,wei12}, optical \citep{car10}, or X-ray bands, as discussed here and in  \citet{eva10,sha10,ten10,fer10,hor10,cus11,ten11,tav11,str11b,wei12}.

\begin{figure}[t!]
\resizebox{\hsize}{!}{\includegraphics[angle=-90,width=6.75cm]{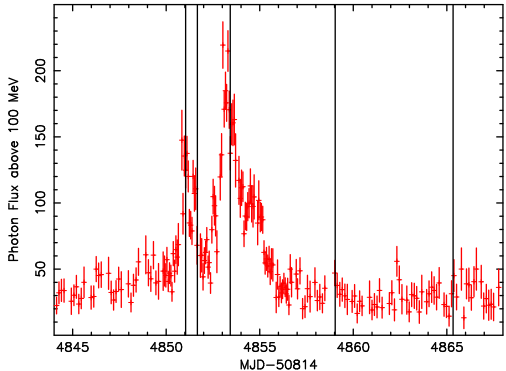}}
\caption{\footnotesize
\Fermi-LAT photon flux ($10^{-7}$~ph/(cm$^2$~s)) above 100~\MeV\ during the 2011-April flare as a function of time. 
Displayed data extend somewhat beyond the time span shown in \citet{bue12} but follow the same data processing as described there.
The vertical lines mark times of the $5$ \cha\ observations discussed in the text.
}
\label{f:lc}
\end{figure}

Figure~\ref{f:lc} shows the \Fermi-LAT light-curve for the 2011-April flare \citep{bue12}. 
For this flare the source doubled its $\gamma$-ray flux within eight hours and reached a peak 30-times the  average.
The (assumed) isotropic luminosity increased to $2\times10^{37}$~{\rm erg/s} in about $10$~hr and the spectrum peaked at $\approx400$~\MeV.
(See \citet{bue12} for details.)
Notification as to the level of flaring prompted us to trigger pre-approved Target-of-Opportunity observations with \cha\ and Figure~\ref{f:lc} also indicates the times of these observations.
\section{The X-ray Observations}
The five observations (ObsIDs $13150-13154$) used the (back-illuminated) ACIS S3 CCD approximately centered on the Crab pulsar, during and somewhat after the 2011-April $\gamma$-ray flare.
For these observations, the spacecraft was dithered with an amplitude set to $1\arcsec$.
Although standard processing typically produces an aspect solution better than $0.5\arcsec$, this small uncertainty can still introduce noticeable shifts amongst different data sets.  
Thus, we re-registered images using the read-out streak and the hole in the images produced by the severely piled-up pulsar.

Owing to the high flux, we used a special mode with 0.2-s frame time, which limits the CCD read-out to a $300\times 300$~ACIS-pixel ($\approx 150\arcsec\times 150\arcsec$) subarray.
Although each observation lasted about 10 ks, telemetry saturation reduced the effective integration time to approximately 1200 s per observation.
Despite the short frame time, regions of high surface brightness suffer somewhat from pile-up effects. 
In view of interstellar absorption at low energies and declining flux at high energies, we limited the analysis to data in the energy range 0.5--8.0~\keV.
We then searched for X-ray variations.

\section{X-ray Image Analysis} 

\begin{figure}[]
\resizebox{\hsize}{!}{\includegraphics[angle=-90,width=6.75cm]{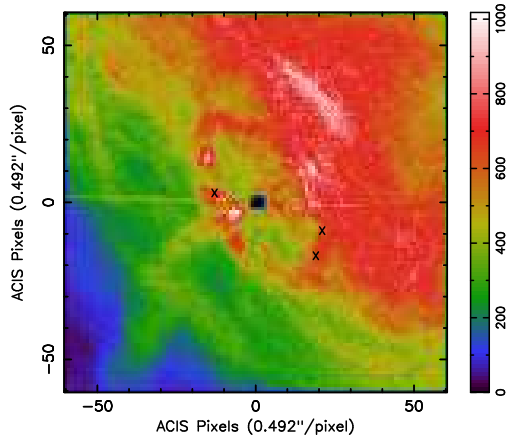}}
\caption{\footnotesize
Summed count image for the 5 \cha~ observations during the $\gamma$-ray flare.
North is up and the pulsar is at (0,0). 
The nearly horizontal streak through the location of the pulsar is the trailed (out-of-time) image, resulting from the very short exposure of each pixel as the image is read out.
As the 5 observations occurred at slightly different roll angles, the read-out streak is slightly blurred. 
The $\mathsf X$ symbols mark locations of the 3 statistically most significant variations, with the most significant being the one to the east of the pulsar.
\label{f:xrayimage}
}
\end{figure}

For each observation, we re-binned a $120\times 120$~ACIS-pixel image centered on the pulsar into a $60\times 60$~array of $2\times 2$~ACIS pixels. 
Each of these $I = 3600$ ``analysis pixels" is sufficiently large (about 1 square arcsec) to enclose most of the \cha~ point spread function anywhere in the field of view. 
Figure~\ref{f:xrayimage} shows the counts per analysis pixel, summed over the 5 observations.

For each analysis pixel $i$, we calculated the mean count rate $r_i$ averaged over the $J = 5$ observations, weighted\footnote{\ $r_i = \sum_{j=1}^{J} \{r_{ij}/\sigma_{ij}^{2}\} / \sum_{j=1}^{J} \{1/\sigma_{ij}^{2}\}$} by the respective (counting-rate) statistical error $\sigma_{ij}$.
For evaluating the statistical significance of temporal variations over the $J = 5$ observations, we compute\footnote{\ $\chi_{i}^{2} = \sum_{j=1}^{J} \{(r_{ij}-r_{i})^{2}/\sigma_{ij}^{2}\}$.} $\chi_{i}^{2}$.
Figure~\ref{f:xrayimage} also indicates the three pixels which showed the highest significance based on $\chi_{i}^{2}$.

The most significant variation has $\chi_{i}^{2} = 23.5$ on $\nu = (J-1) = 4$ degrees of freedom.
Such a fluctuation is expected statistically in at least 1 of 3600 pixels in 31\% of realizations and thus is not considered terribly significant. 
Based upon the $\chi^{2}$ probability distribution and the number of ``tries'', a 99\%-confidence detection would require $\chi_{i, 99\%}^{2} > 31.2$ on $(J-1) = 4$~degrees of freedom.
While we detect no statistically significant variations at 99\% confidence, it is perhaps curious that the 3 most significant variations occur at locations on the inner ring. 
 
Other effects, such as changes in the roll angle of the read-out streak, can lead to possibly spurious variability. 
This may be the case for the analysis pixel with the most significant variation, which lies to the east of the pulsar but adjacent to the average read-out streak (Figure~\ref{f:xrayimage}).

\subsection{Limits to the X-ray Flux}

Neglecting for the moment the effects of pile-up, the photon spectral flux is proportional to the count rate for an assumed spectral shape.
Consequently, any change in count rate corresponds to a proportionate change in the photon spectral flux.
Using the \cha~ PIMMS\footnote{http://asc.harvard.edu/toolkit/pimms.jsp} for ACIS-S and an absorption column $N_{\rm H} = 3.1\times 10^{21}\ {\rm cm}^{-2}$, we determine this constant of proportionality for an X-ray power-law photon index $\Gamma_{x} = \frac{2}{3}$, 1, and 2:
At $E_{x} = 1$~\keV, $N_{E}(E_{x})/r =$\ 0.99, 1.26, and 2.46 $\times 10^{-3}$~ph/(cm$^2$ s \keV) per ct/s, respectively.
Table~\ref{t:r2} shows our calculations of the upper limits to the photon spectral flux $N_{E}(E_{x})$, the energy spectral flux $F_E(E_{x})$, and the indicative (isotropic) luminosity $E L_{E}(E_{x}) = 4 \pi D^{2} E F_{E}(E_{x})$ at $D = 2$~kpc, for the analysis pixel with the most significant X-ray variation.
Correcting for pile-up has little effect in low-count-rate regions, but would raise these flux upper limits by $10$\% or so for the high-count-rate regions.

\begin{table}
\begin{center}
{\caption {99\%-confidence upper limits for various parameters at 1~\keV\ for the analysis pixel with the most significant variation. \label{t:r2}}}
\begin{tabular}{cccc} \hline
$\Gamma_{x}$          & $\frac{2}{3}$ & 1      & 2  \\ \hline
$N_{E}~^a$            & 0.55          & 0.70   & 1.36  \\
$F_{E}~^b$            & 0.88          & 1.12   & 2.18  \\
$E L_{E}~^c$          & 0.42          & 0.54   & 1.05  \\
$\Gamma_{x\gamma}$    & 1.20          & 1.22   & 1.27  \\ \hline
\end{tabular} 
\end{center}
$^a$  $10^{-4}$ ph/(cm$^2$ s \keV) \\
$^b$  $10^{-13}$ erg/(cm$^2$ s \keV) \\
$^c$  $10^{32}$ erg/s 
\end{table}

\subsection{Constraints on the X-ray to $\gamma$-ray Spectral Index}

We can also compare the $\gamma$-ray data to the X-ray data to quantify the implications of our lack of detection of time variations in the latter. 
The approach compares a variability measure for the X-ray (1-\keV) photon spectral flux $\Delta N_{E}(E_{x})$ in each analysis pixel with the analogous variability measure for the $\gamma$-ray (100-\MeV) photon spectral flux $\Delta N_{E}(E_{\gamma})$.
Specifically, we calculate the sample standard deviation of the $\gamma$-ray spectral flux at 100~\MeV~ using power-law fits to the 5 \Fermi-LAT measurements that were simultaneous with the 5 \cha~ observations.
For these 5 observations, the mean and sample standard deviation of the photon spectral flux at 100~\MeV\ are $1.21\times 10^{-10}$ and $5.77\times 10^{-11}$ ph/(cm$^2$ s \keV), respectively.

Based upon the sample standard deviation of the photon spectral flux at $E_{x} = 1$~\keV\ for each X-ray analysis pixel and the measured standard deviation, $5.77\times 10^{-11}$ ph/(cm$^2$ s \keV) at $E_{\gamma} = 100$~\MeV, we constrain the effective X-ray to $\gamma$-ray photon index of the flaring component: $\Gamma_{x\gamma} \equiv -\log[\Delta N_{E}(E_{\gamma})/\Delta N_{E}(E_{x})] / \log[E_{\gamma}/E_{x}]$. 

In that the $\gamma$-ray variations are statistically significant and the X-ray variations are not, we compute 99\%-confidence upper limits to $\Gamma_{x\gamma}$ (Table~\ref{t:r2} last row).
The $99$\%-confidence limits to $\Gamma_{x\gamma}$ are marginally consistent with the low-energy extrapolation of the $\gamma$-ray spectrum ($\Gamma_{\gamma} = 1.27 \pm 0.12$) of the flaring component \citep{bue12}.

\section{Conclusions}
Using \cha, we acquired X-ray images of the Crab Nebula contemporaneous with the 2011-April $\gamma$-ray flare. 
We tested for time variations amongst the 5 pointings, each with an effective exposure time $\approx 1200$~s and a minimum separation of 0.6~days. 
We did not detect statistically significant X-ray variations; thus we can set only upper limits to any X-ray variations associated with the $\gamma$-ray flare.
As the \cha~ ACIS images suffer severe pile-up near the Crab pulsar, our search for variability in the X-ray images was not sensitive to variations within the central $\approx 1.5\arcsec$ or so.

Comparing the upper limits to X-ray variations with the \Fermi-LAT-measured $\gamma$-ray variations, we set upper limits at $99$\%-confidence to the effective X-ray--$\gamma$-ray photon power-law index $\Gamma_{x\gamma} \leq 1.20$ to $ \leq 1.27$, dependent upon assumptions about the X-ray index $\Gamma_{x}$.
As \Fermi-LAT measures a $\gamma$-ray index $\Gamma_{\gamma} = 1.27\pm 0.12$ for the flaring component, it is statistically possible that the flaring component's spectrum extends as a simple power-law from $\gamma$-rays to X-rays.
Further, we note that our upper limit to $\Gamma_{x\gamma}$ is consistent with transparent synchrotron emission, whose photon index must be $>\frac{2}{3}$.

Elsewhere \citep{wei12} we present a more detailed analysis of the X-ray data including the results of searches for variability within each observation. 
\citet{wei12} also discusses a Keck near-IR observation of the inner knot ($\approx 0.65\arcsec$ from the pulsar) made in conjunction with the 2011-April flare and a number of VLA observations searching for a point source appearing either at an unusual location and/or contemporaneous with the flare. 

Although no ``smoking gun" has been identified, we are encouraged that we have identified a number of regions in the X-ray images that are possible candidates. 
In addition we (see acknowledgments) have also established further Target of Opportunity observations with \cha\ and HST that will be triggered at the onset of the next $\gamma$-ray flare. 
The X-ray observations will also probe the region very close to the pulsar using the \cha~ High-Resolution Camera (HRC). 

\begin{acknowledgements}
I am very grateful to all of my colleagues participating in this collaboration: 
Allyn~F.~Tennant, Jonathan Arons, Roger Blandford, Rolf Buehler, Patrizia Caraveo, C.~C.~Teddy Cheung, Enrico Costa, Andrea de Luca, Carlo Ferrigno, Hai Fu,
Stefan Funk, Moritz Habermehl, Dieter Horns, Justin~D.~Linford, Andrei Lobanov, Claire Max, Roberto Mignani, Stephen~L.~O'Dell, Roger W. Romani,
Edoardo Striani, Marco Tavani, Gregory~B.~Taylor, Yasunobu Uchiyama, \& Yajie Yuan.
\end{acknowledgements}

\bibliographystyle{aa}

\end{document}